# Behavioral individuality reveals genetic control of phenotypic variability

**Julien F. Ayroles[1,2,3, †], Sean M. Buchanan[4], Chelsea Jenney [1,4,5], Kyobi Skutt-Kakaria[1,5],**
**Jennifer Grenier[3], Andrew G. Clark[3], Daniel L. Hartl[1], Benjamin L. de Bivort[1,4,5, †]**

1 – Dept. of Organismic and Evolutionary Biology, Harvard University, Cambridge, MA 02138, USA.
2 – Harvard Society of Fellows, Harvard University, Cambridge, MA 02138, USA.
3 – Dept. of Molecular Biology and Genetics, Cornell University, Ithaca, NY 14853, USA.
4 – Rowland Institute at Harvard, Cambridge, MA 02142, USA.
5 – Center for Brain Science, Harvard University, Cambridge, MA 02138, USA.

[†] Corresponding authors: ayroles@fas.harvard.edu, debivort@oeb.harvard.edu

**Abstract**

Variability is ubiquitous in nature and a fundamental feature of complex systems. Few studies, however, have investigated variance itself as a trait under genetic control. By focusing primarily on trait means and ignoring the effect of alternative alleles on trait variability, we may be missing an important axis of genetic variation contributing to phenotypic differences among individuals[1,2]. To study genetic effects on individual-to-individual phenotypic variability (or intragenotypic variability), we used a panel of *Drosophila* inbred lines[3] and focused on locomotor handedness[4], in an assay optimized to measure variability. We discovered that some lines had consistently high levels of intragenotypic variability among individuals while others had low levels. We demonstrate that the degree of variability is itself heritable. Using a genome-wide association study (GWAS) for the degree of intragenotypic variability as the phenotype across lines, we identified several genes expressed in the brain that affect variability in handedness without affecting the mean. One of these genes, *Ten-a* implicated a neuropil in the central complex[5] of the fly brain as influencing the magnitude of behavioral variability, a brain region involved in sensory integration and locomotor coordination[6]. We have validated these results using genetic deficiencies, null alleles, and inducible RNAi transgenes. This study reveals the constellation of phenotypes that can arise from a single genotype and it shows that different genetic backgrounds differ dramatically in their propensity for phenotypic variability. Because traditional mean-focused GWASs ignore the contribution of variability to overall phenotypic variation, current methods may miss important links between genotype and phenotype.

**Abbreviations:** DGRP: Drosophila Genome Reference Panel; ANOMV: Analysis of means for variance; QTL: quantitative trait loci, GWAS: genome wide association study, MAD: median absolute deviation. GWAS: genome wide association study, CI: confidence interval.

Quantitative genetics was founded on a concept in which phenotypic variation is explained solely by differences in mean phenotypes among genotypes. Under this models, intragenotypic variability is assumed to be attributable to non-genetic environmental perturbations[1]. There is however, growing evidence for the importance of genetic control of variance[2,7,8] and that variance itself is a quantitative trait. Although studies of morphology[9-11] and animal breeding[12,13] have long noted the heterogeneity of variance among genotypes, this axis of variation has received little attention compared to the effect of genetic variation on trait means. As a result, the mechanisms by which variable phenotypes arise from a uniform genetic background are still poorly understood, particularly in the context of behavior, where variability may be a critical determinant[14,15]. Most recently with the advent of genome-wide association studies, several groups[7,8,16,17] have mapped quantitative traits loci affecting variance (vQTLs) by comparing phenotypic variances among individuals that share genotypes. These studies examine the average effect of QTL alleles across genetic backgrounds and heterogeneous environmental across individuals[18], in the process losing any specific effects intrinsic to each individual.

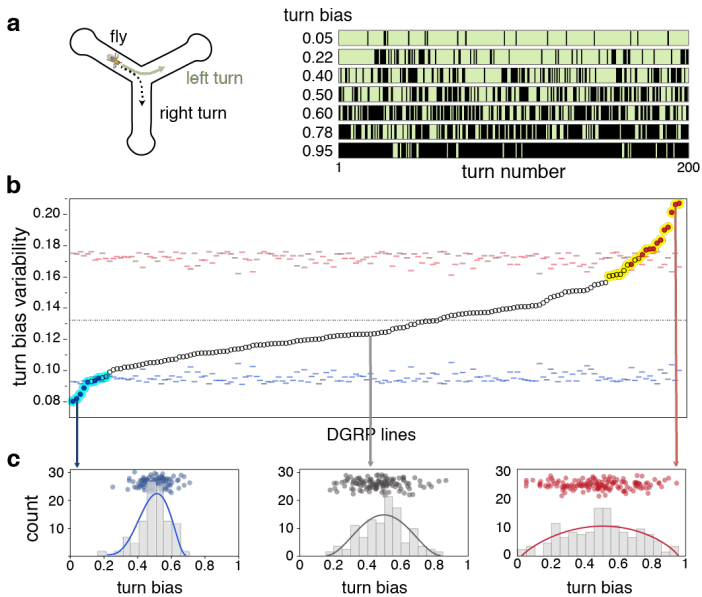

**Figure 1 – Intragenotypic variability of locomotor handedness varies across DGRP lines.** *a.* Diagram of the Y-maze used to quantify individual locomotor behavior. Plot at right illustrates 200 sequential turns for 7 representative individual flies. A turn bias of 0.05 indicates that this particular fly turned right 5% of the time (black stripes indicate right turns and green stripes left turns). *b.* Sorted distribution of the standard deviations of within-line individual turn bias, for 159 DGRP lines. Red and blue filled dots are significant, exceeding their corresponding tick-marked 99.9% confidence intervals, estimated by permutation. See table S1 for experimental sample sizes. Cyan and yellow highlighted dots are significant at $p < 0.001$ based on non-parametric bootstrap. *c.* Distributions of turning bias across individuals for three representative DGRP lines with low, intermediate and high intragenotypic variability. Each dot represents the turning bias of a single fly within that line. Lines are beta distribution fits, chosen because they model over-dispersed binomial distributions.

Here, we examine diversity that is typically hidden in population averages, by examining phenotypic variability among individuals with the same genotype. This is the variation that we would observe if we could generate a large number of copies of individuals of the same genotype in a common environment, and measure a trait across them (an experiment for which isogenic lines[2,9-11,18] are especially suited). In this case phenotypic differences among genetically identical individuals result from subtle micro-environmental perturbations and stochasticity in development, while differences in variability among genotypes reflects genetic differences in developmental stability[11]. While intragenotypic variability contributes to phenotypic variation in a population, this source of variation is not usually estimable because, with few exceptions, each individual in an outbred diploid population possesses a unique instance of its genotype. As a consequence we have little understanding of the causes and consequences of inter-individual intragenotypic variability. It nevertheless has wide ranging implications. In evolutionary biology, variability offers an adaptive solution to environmental changes[19,20]. In medical genetics, diseased states emerge beyond a phenotypic threshold, and high variability genotypes will produce a larger proportion of individuals exceeding that threshold than low variability genotypes, even if they have the same mean. While intragenotypic variability has been discussed in animal behavior, particularly in the context of the emergence of personality[14,21], to date no genes have been associated with behavioral variability that do not also affect the mean.

To study this problem, we used a panel of *Drosophila* inbred lines where, for each line, we can phenotype a large number of individuals of the same genetic background, age, and rearing environment. This allowed us to empirically estimate the magnitude of intragenotypic variability. Specifically, we measured the locomotor behavior of flies walking freely in Y-shaped mazes[3], focusing on the variability in locomotor handedness (left-right turning bias). The precision and high-throughput nature of our assays allows a large number of flies to be measured per genotype and permits robust estimates of the sampling error on variance itself. We tracked two hours of locomotor behavior of 110 individuals (on average) from each of 159 lines from the *Drosophila* Genetic Reference Panel (DGRP) in a randomized block design. For each individual fly, we recorded the time and left-right direction of each turn in the maze (**Fig 1a**), estimating a turn bias score as the fraction of turns that were to the right. Flies performing more than 50 turns were analyzed, and completed 413 turns per trial on average.

We began by comparing the mean turning bias and found no significant genetic variation across lines (**Supplementary Fig 1**). In other words, averaged across individuals within a line, each lines are unhanded, making an equal proportion of left and right turns. We verified the lack of genetic variation for turning bias *within lines* by crossing pairs of males and females with matched turning biases (*e.g.*, two strongly right-biased parents). For all crosses, the phenotypic mean and variance of the distribution of the $F_1$ generation was statistically indistinguishable from the distribution of the parental line (**Supplementary Fig 2**). Handedness therefore provides an ideal framework to study the genetics of variability because genetic effects on variability are not confounded by mean effects. Next, using parametric (ANOVA) and non-parametric (bootstrapping) statistical approaches, we compared levels of intragenotypic variability across lines and found highly significant among-line differences in variability, implying that the abundance of individuals that were either strongly left- or right-biased was itself variable among lines. This indicates that the degree of intragenotypic variability itself is under genetic control in these lines (**Fig 1b, Supplementary table 1**). In order to obtain further evidence that intragenotypic variability is heritable, we mated two high-variance and two low-variance lines to each other and measured turning bias in the resulting progeny (phenotyping an average of 183 individuals per cross). Intercrosses between high-variance lines led to high variance F1 progeny and crosses with low-variance lines yielded low variance F1 progeny **(Fig 2, Supplementary table 1)**. In both cases the variability in the F1 progenies was statistically indistinguishable from that of the parents.

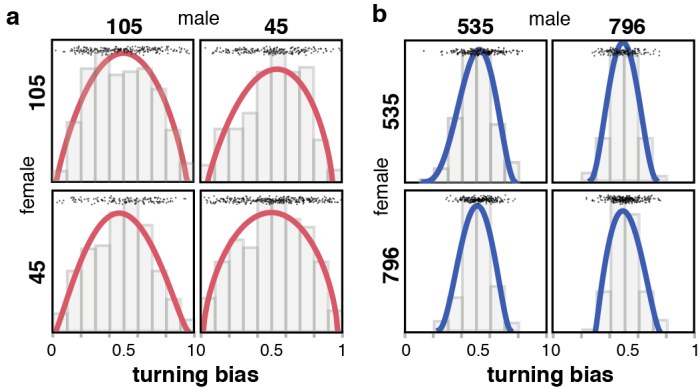

**Figure 2 – Intragenotypic variability for turning bias is heritable.** *a.* Distribution of F1 turn biases resulting from high variance line 105 reciprocally crossed to high variance line 45 (Brown-Forsythe $p = 0.08$; $n_{105x105} = 235$; $n_{45x45} = 315$; $n_{105*45} = 223$; $n_{45*105} = 135$). *b.* Distribution of F1 turn biases resulting from low variance line 535 reciprocally crossed to low line variance line 796 (Brown-Forsythe $p = 0.02$; $n_{535x535} = 197$; $n_{796x796} = 265$; $n_{796*535} = 160$; $n_{535*796} = 234$). In both panels, the progeny are presented on the off diagonal. Lines are beta distribution fits. Points are individual flies. For both 2a. and 2b., *p*-values comparing F1 to parents ranged form 0.14 to 0.99, uncorrected for multiple comparisons.

It is conceivable that some lines might be better than others at buffering microenvironmental perturbations, in which case the degree of intragenotypic variability among lines would be correlated across traits. To test this possibility, we scored additional phenotypes from our Y-maze data, namely, the total number of turns (a measure of overall activity); the mutual left-right information between successive turns; and the regularity of turn timing. We also analyzed other phenotypes previously measured on the DGRP (starvation resistance[22], chill coma recovery[22], startle response[22], and night sleep[23]). We found significant genetic variation for variability in all these phenotypes, confirming that genetic control of variability is ubiquitous across phenotypes. On the other hand, we found no evidence that the variances of these traits are correlated across phenotypes (with the sole exception of mean absolute deviation (MAD) of turn bias and turn timing regularity) **(Supplementary Fig 3)**. This suggests that the genetic basis for intragenotypic variability is trait-specific (and implicates many independent loci controlling this often-ignored trait).

The DGRP lines have been fully sequenced[22], allowing for whole-genome association mapping using the variability (i.e. MAD) of turning bias as a trait. While the DGRP is underpowered to study the architecture of complex traits due to the relatively small number of lines (*n*=159 in this study), it is a good resource to identify candidate genes for experimental follow up[3,24]. To that end, we performed an association study using a series of locus-specific mixed linear models (accounting for relatedness between lines as well as experimental block effect) and found 36 polymorphisms in 22 genes associated with variability in turning bias using a nominal *p*-value[3,22] of $5 \times 10^{-6}$ **(Supplementary table 2)**. These genes are strongly enriched for expression in the central nervous system both in adults and in larvae (adult CNS enrichment in adult FET $p < 0.001$ and in larvae FET $p < 0.01$, data from FlyAtlas[25]) **(Supplementary Fig 4)**. Among these, the synaptic target recognition gene *Tenascin accessory* (*Ten-a,* GWAS $p < 3 \times 10^{-6}$) **(Fig 3a)** caught our attention, *Ten-a* is a transmembrane signaling protein involved in synapse formation[26,27], critical to the development of the brain central complex[5] (a brain structure implicated in sensory integration and locomotion[4,6,28]) and is highly conserved from insects to mammals[29]. In order to validate the role of *Ten-a* in modulating variability in turning bias, we used a null allele (*Ten-a[cbd-KS9635]*), a deficiency overlapping *Ten-a* (*Df(1)Ten-a[26]*), and expression knock-down using RNAi (UAS-TRiP.JF03375[30]). In all cases, disrupting *Ten-a* increases the variability in turning bias with no effect on the mean **(Figs 3b,c and 4)**. The effect of RNAi knock-down suggests a quantitative relationship between *Ten-a* mean expression and variance in turning bias.

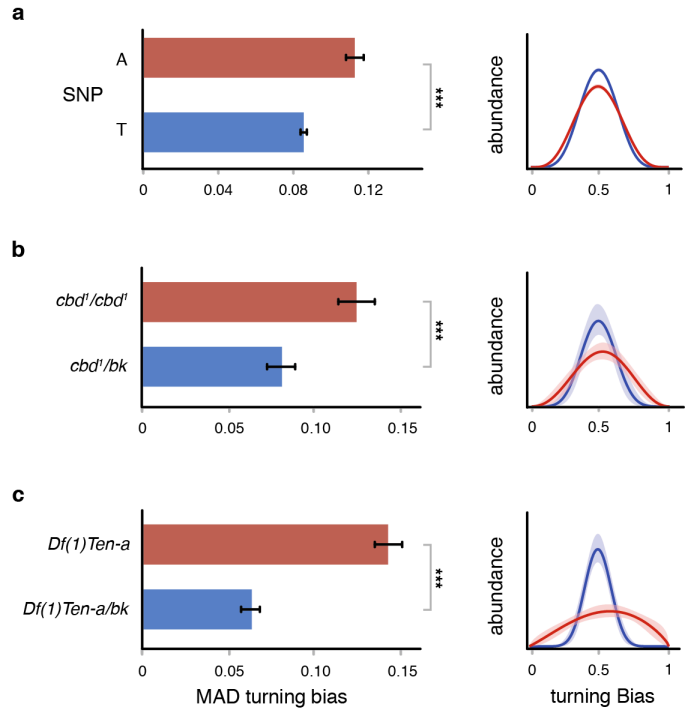

**Figure 3 – Effect of *Ten-a* mutation on intragenotypic variability.** *a.* Intragenotypic variability (MAD) in turn bias of flies harboring alternative alleles of the *Ten-a* SNP identified in our GWAS ($n = 159$, GWAS $p < 3 \times 10^{-6}$). *b.* Turn bias MAD of a homozygous *Ten-a* null allele (*cbd[1]*; red) and heterozygous control (blue). *bk* indicates the *Ten-a[+]* genetic background Berlin-K. $n_{cbd1/Bk} = 59$, $n_{cbd1/cbd1} = 99$, Brown-Forsythe $p = 0.0074$, bootstrapping $p < 0.001$. *c.* Turn bias MAD of a line bearing a homozygous deficiency overlapping *Ten-a* (red) and heterozygous control (blue). $n_{Df(1)-Bk} = 100$, $n_{Df(1)Ten-a} = 97$, Brown-Forsythe $p = 1.5 \cdot^{11}$, bootstrapping $p < 0.001$. *** $p < 0.001$. Right plots in all panels are corresponding beta distribution fits of the distribution of turn bias scores within each experimental group. Shaded regions are 95% CIs on the beta fits, estimated by bootstrap resampling; CIs in (*a*) small compared to line thickness. Error bars are +/- one standard error estimated by bootstrap resampling.

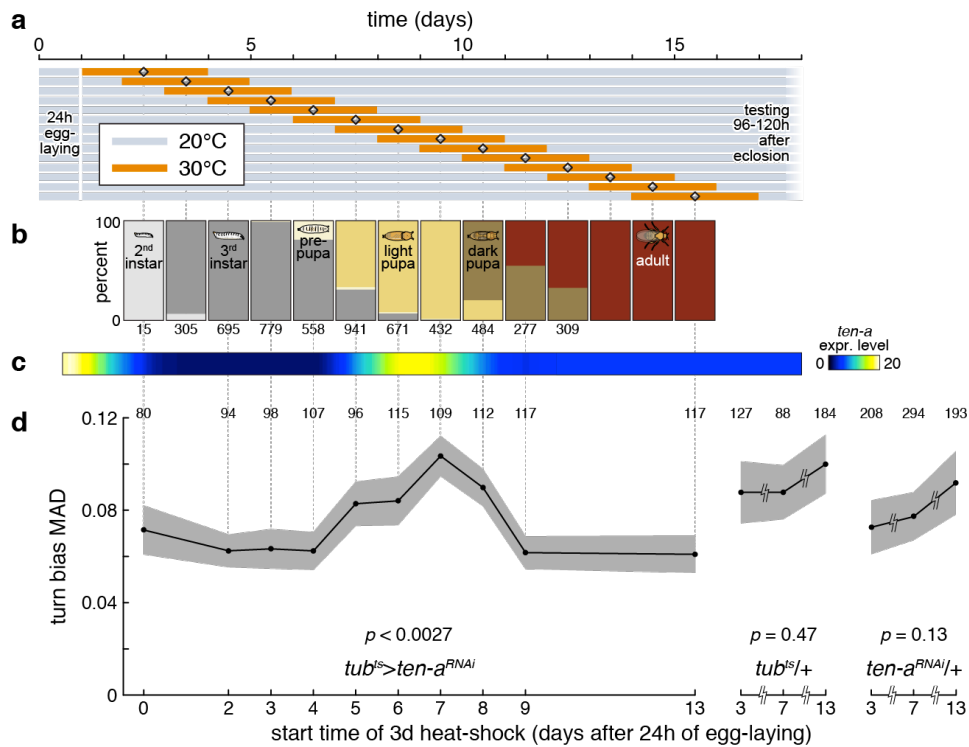

**Figure 4 – Disruption of *Ten-a* expression in mid-pupa affect behavioral variance.** *a.* Time-courses of sliding window *Ten-a* RNAi induction. Flies laid eggs for 24 h prior the start of the experiment and were reared at 20°C (gray) until three days of RNAi induction at 30°C (orange). Flies were then returned to 20°C until they were tested 3 to 5 days post-eclosion. *b.* Fraction of flies at any developmental stage during the course of the experiment. Numbers indicate sample sizes. *c. Ten-a* expression level over development. Expression level derived from modENCODE[22]. *d.* Effect of temperature inducible *Ten-a* RNAi on the variability of turning bias over development. Knock down effect varied significantly with the timing of the induction window ($p = 0.0027$) estimated by a bootstrapping omnibus test (see Methods), with a knock down starting on day 7 greatly increasing variability. This knockdown window coincides with the peak of *Ten-a* expression during pupation. Grey regions represent +/- standard error, estimated by bootstrapping. To the right, the controls, *tub[ts]/+* and *Ten-a[RNAi]/+*, measured after 3 day 30°C windows starting on days 3, 7 and 13, show no effect ($p < 0.47$ and $p < 0.13$ respectively). Numbers above data indicate sample sizes. Vertical guide lines associate data points across panels.

The bias in handedness of a given fly is a fixed property of that individual (*e.g.* a young adult with a strong left bias will display this bias throughout its life[4]). This suggests that handedness may be wired during development. To determine if there is a critical developmental period when *Ten-a* expression is required to regulate variability, we used temperature-inducible RNAi to knock down *Ten-a* in sliding 3-day windows (**Fig 4a and b**). We found that knocking down *Ten-a* expression in mid-pupae increases the resulting adults' variability (**Fig 4d**). This stage coincides with a spike in *Ten-a* expression[31,32] (**Fig 4c**) and the formation of the central complex[5]. Further, Buchanan *et al*[4] mapped a set of neurons within the central complex (i.e. protocerbral bridge columnar neurons) that regulates the magnitude of left-right turn bias, and therefore the magnitude of intragenotypic variability.

Our use of inbred lines allowed us to estimate a parameter (intragenotypic variability) that otherwise could not be observed and uncovers the spectrum of phenotypes a given genotype can produce. Because all phenotypic variation in handedness in a population is (surprisingly) attributable to intragenotypic variability, it highlights the importance of genetic control of variability. The significance to medical genetics is far-reaching, specifically for attempts to predict phenotype from genotypes (e.g., "missing heritability[33]"). Within evolutionary biology, this is significant for understanding evolutionary trajectories or forces maintaining polymorphisms[9,19,34]. An individual drawn from a high-variability genotype has the potential to explore a wider range of phenotypic space than one with a low-variability genotype (far beyond what may be determined by the mean effect alone, **Supplementary Fig 5**). This could be advantageous in evolutionary adaptation, but in human genetics it could be deleterious when an extreme phenotype enhances disease risk. Recent progress in single-cell technology18 now makes it possible to measure cell to cell variation within an individual genotype, enabling the study of intragenotypic variability in human phenotypes such as drug response and cancer progression. This study adds to an emerging body of work suggesting that in order to understand variation in complex traits, we must also understand the contribution of intragenotypic variability.

## Methods

All raw data, data acquisition software and analysis scripts are available at *http://lab.debivort.org/genetic-control-of-phenotypic-variability.*

### Drosophila stocks

The DGRP was created as a community resource for the genetics mapping of complex traits[3]. It consists of a collection of isofemale lines derived from a single field collection from the Raleigh NC farmers market, followed by 20 generations of full-sib mating that rendered most loci homozygous within lines (expected F= 0.986[22]). As a result, the genetic variation that was present between individual flies in the natural population is now captured between lines in the panel. This allows us to measure any phenotype on a given genotypic background and phenotype the same genotype a large number of times in any environment. Completion of the genome sequencing for all lines combined with *Drosophila*'s generally rapid decay in linkage disequilibrium between polymorphic sites makes the DGRP a powerful tool to identify genetic polymorphisms that affect quantitative phenotypes[3]. The DGRP lines are available from the Drosophila Stock Center (http://flystocks.bio.indiana.edu). We used a total of 159 lines in this study (lines with the highest inbreeding coefficient - list is provided along with data at: http://lab.debivort.org/genetic-control-of-phenotypic-variability. Stock used for *Ten-a* line: Berlin-K, *central-body-defect*[KS96 28], *Df1-Ten-a*[5], and RNAi TRiP.JF03375[30]

All flies were reared on standard fly media (Scientiis and Harvard University BioLabs fly food facility), in a single 25°C incubator at 30-40% relative humidity with a 12/12h light/dark cycle. Before each assay flies were fully randomized across: blocks, lines, Y-maze arrays and position on the array. At least 3 strains were assayed simultaneous on each array.

### Variance, variation, variability:

The similarity between concepts of variance, variation and variability may lead to some confusion. The meanings of these terms are reviewed in Wagner and Altenberg[40]. In accordence with their definition, we used the terms variance[2,18] to describe the standard statistical dispersion parameter ($\sigma^2$) or estimates of it derived from observations ($s^2$). Variability refers to the potential of an organism or genotype to vary phenotypically. Variation refers to the realized (observable) differences between individuals or genotypes.

### Phenotypic assay

Studying variance as a trait poses a number of challenges including: the large sample size required (precise estimates of variance requires a larger number of observations than needed to estimate means), the experimental design (as to not confound sources of error), and potential measurement error of the phenotype itself[18]. It is with these considerations in mind that we developed a high throughput assay aimed at monitoring the behavior of individual flies placed into individual Y-mazes[4] (Fig 1.a). Each experiment examines one array of 120 Y-mazes (refered to as maze-arrray). Mazes were illuminated from below with white LEDs (5500K, LuminousFilm), imaged with 2MP digital cameras (Logitech), and the X-Y positions of each fly' centroids were automatically tracked and recorded with software custom written in LabView (National Instruments). Further details about the assay are provided in[4], code available at *http://lab.debivort.org/neuronal-control-of- locomotor-handedness/.*

While various statistics can be computed to estimate the degree of variability of a distribution, in this study we use one the most robust metrics, the Median Absolute Deviation (MAD)[8,17]. It is defined as the median of the absolute deviation from each observation's median: MAD = median ($|X_i - median(X_i)|$). Where $X_i$ is the phenotypic score of an individual fly within a line. MAD scores were computed for each line for each phenotype.

Only females were used in this experiment and only lines yielding data from a minimum of 75 individuals were included. Before each assay flies were very lightly anesthetized, rapidly transferred to an individual Y-maze and given a recovery period of 20 minutes before the start of the assay. Fly behavior in the mazes was monitored for two hours. This assay generated four phenotypes: (1) The handedness or left/right turning bias in the arms of the maze summed over all L/R decisions. A turning bias score of 0.8 for a given fly would indicate that this individual made left turns 80% of the time at the maze's junction over the two hour period. This simple phenotype is particularly well suited for this study given that it is measured without error and the high number turns for any given fly ensure a robust estimate of the turning bias and it variance for each fly. (2) The number of turns over the 2 hour period, an estimate of overall locomotor activity. (3) The "switchiness" or the mutual left-right information between successive turns right/left turn sequence (e.g. LLLLLRRRRR: low switchiness, high mutual information; LLRLLRRRLR: moderate switchiness, low mutual information; LRLRLRLRLR: high switchiness, high mutual information) defined as:

$$(N_{(L,R)}+N_{(R,L)})/(2N_RN_L/N)$$

where $N_{(L,R)}$ is the number of left turns followed by right turns, $N_{(R,L)}$ is the number of right turns followed by left turns $N_R$ is the number of right turns, $N_L$ is the number of left turns, and $N$ is the total number of turns. (4) The regularity of turn timing: a fly with a high score makes turns uniformly throughout the experiment while a low score would characterize a fly making a small number of dense streaks of turns but is inactive for dozens of minutes at a time. It is defined as MAD(ITIs)/(7200/$N$) where ITIs is the vector of inter-turn intervals in seconds. The left/right turning bias is the main focus of this study, additional traits were measured to illustrate that the degree of variability across traits is not correlated between lines.

### Quantitative genetic analysis

*Analysis of Means:* In order to determine if there was genetic variation segregating in the DGRP affecting the mean turning bias, we partitioned the variance for line means using the ANOVA model:

Y = μ + L$_{random}$+ B$_{random}$ + L*B $_{random}$ + A + X + A*X + $e$

where Y is turning bias score of each fly; *L* is the effect of line treated as random B is the effect of block treated as random, X the box effect, A the maze-array effect and *e* is the error variance (Supplementary Table 1). ANOVA implemented using PROC MIXED in SAS 9.3.

*Variance heterogeneity:* We used several statistical approaches to estimate heterogeneity of variance for turning bias between lines (Supplementary Table 1). (1) The Brown-Forsythe test which is based on a one way ANOVA and relies the absolute deviation from the median[8,36]. (2) Non parametric bootstrapping in which we first pooled all the turn bias scores for all individual flies across lines, then resampled each line experimental group from this pool, matching the sample size. Lines in which the MAD of the resampled group was closer to the MAD of the pooled data, in fewer than 10 of 10,000 resamples, were taken as significant. This tests the null hypothesis that each group is drawn from an identical distribution of observations, using MAD as a test statistic. (3) A nonparametric version of the Analysis of Mean for Variances (ANOMV)[37,38]. This approach compares the group means of the Median absolute deviation MAD to the overall mean MAD under the null hypotheses that the group MAD means equals each line specific MAD. Implemented in SAS 9.3[37]. (4) Finally we used the same ANOVA model described above for the analysis of mean but used the absolute deviation from the median[8,9] as a measure for each fly as the dependent variable. Implemented using PROC MIXED in SAS 9.3.

*Phenotypic correlation between traits:* We assesed 4 traits as measured in this study and 4 additional traits gathered from the literature (standard deviation for starvation, startle response, chillcoma recovery[3], coefficient of environmental variation for night sleep[23]). Phenotypic correlation between traits was computed as the pearson product-moment correlation. *P*-values are not corrected for multiple comparison. Implemented using PROC GLM in SAS 9.3.

**High and low variance lines intercrosses**

In order to confirm that variability was heritable, we crossed respectively high and low variability lines to each other. We crossed high variability lines 45 to lines 105 and low variability lines 796 to 535. 10 females and 5 males were used for each cross. Flies were reared and phenotyped using the same protocoled described above. Note that parental behavior was re-measured concurrently with F1 behavior following a corresponding self-cross (e.g. 45 x 45). We assessed statistical significance between parental lines and their progeny using the Brown-Forsythe test and a bootstrapping two-tailed *z*-test (with *n* = 10,000 resamples). We resampled the turn bias of the parents and for each iteration calculated the MAD of turning bias then compared the MAD for the F1 progeny to their parents.

**Genome wide association mapping**

GWAS was performed using the code and approach decribed in[22] (dgrp2.gnets.ncsu.edu). In a first step phenotypic stores were adjusted for the potential effect of Wolbachia and known large inversions segregating in this panel (namely: *In(2L)t, In(2R)NS, In(3R)P, In(3R)K,* and *In(3R)Mo*) none of them were associated

with variability turning bias. We then fitted a series of loci specific mixed linear model using the model:

$$Y = \mu + Sb + Iu + e$$

Where Y is the MAD of turning bias of each DGRP lines, S is the design matrix for the fixed SNP effect b, I is the incidence matrix for the random polygenic effect u, and e is the residual[22]. A total of 1,931,250 SNPs and indels were used in this analyses with the minor alleles present in at least 7 DGRP lines, using only biallelic sites. Polymorphisms segregating within lines were discarded and for each SNP a least 60 DGRP lines had to have been genotyped to be analyzed. Given the number lines available in the DGRP, GWAS will generally be underpowered[24], however our goal is not to describe the overall genetic architecture of each of these phenotypes, rather we seek to identify interesting candidate genes that would provide some insight into the genetic basis of variance control. For this reason we used a liberal threshold of *p* < 10⁻⁶.

The analysis for tissue enrichment was based on FlyAtlas data which are publically available[39]. For each tissue, we used FlyAtlas AffyCalls[25] to determine which genes were expressed in which tissue (using a conservative filter of four out four present calls). To determine significance we used a Fisher exact test comparing the expected number of gene expressed in each tissue across the entire genome to the observed number of gene expressed in each tissue in our gene list.

**Validation of *Ten-a* effect on variability:**

*Ten-a null and deficency*: The turning bias and MAD turning bias of homozygotes of both the null allele *Ten-a*^cbd-KS963 5 and deficiency overlapping *Ten-a Df(1)Ten-a*⁶ where compared to heterozygous animals over their genetic background, Berlin-K.

*Time course knockdown of Ten-a RNAi*: Ten adult *Ptub-Gal80ts;Ptub-Gal4/Sb* females were crossed to three *UAS-Ten-a RNAi y1,v1;P(TRiP.JF03375)attP2* males for RNAi induction. Flies were allowed to mate for 24 hours at 20°C at which point the parents were passaged out and the bottles containing F1 eggs were returned to 20°C until the beginning of their heat shock window. Flies were exposed for 72 hours to 30°C temperature, in 72 hour sliding window each day over 14 windows (fig 4a). All flies assayed were between three and five days post-eclosion. In parallel, each day developing flies of the same genotype were examined and counted to determine the fraction of flies in each developmental stage at the time of RNAi induction (fig 4b). Stages containing larval animals were microwaved to melt the media, poured through a sieve and larval carcasses counted under a dissecting scope. Controls were performed using *Ptub-Gal80ts;Ptub-Gal4/Sb* females crossed to Canton-S males and Canton-S females crossed to *UAS-Ten-a RNAi y1,v1;P(TRiP.JF03375)attP2* males (fig 4d), otherwise treated identically.

*Ten-a expression*: Data for *Ten*-a expression over developmental time (Fig 4c) were downloaded from FlyBase[32] and derived from ModEncode[31] (modENCODE DDC ids: modENCODE_4433, _4435 and _4439 through _4462). This data reflects animals synchronized by developmental stage to within two hours. To make this data comparable to our experimental groups, in which egg-laying occurred over 24h, we corresponded the developmental stages of the FlyBase data to our developmental stage time course (fig 4b), linearly interpolated the expression values and applied a 24h sliding window average to the interpolated data, mimicking the dispersion effects of our longer egg-collection window.

## Conflict of interest

The authors have no financial interests related to this work.

## Author Contributions

JFA and BLdB designed the experiments, JFA, SMB, CJ, BLdB, JG, and KSK carried out the experiments, JFA and BLdB analysed the data. JFA, SMB, AGC, BLdB, DLH and KSK, wrote the manuscript.

## Acknowledgements

We are grateful to Ian Dworkin, Mia Levene, Noah Zaitlen, and Eric Stone for comments, discussions and helpful feedback on the manuscript. This work was supported by grants: NIH R01 AI064950 to A.G.C., Harvard Society of Fellows Fellowship and Harvard Milton Funds to J.F.A, and the Rowland Junior Fellowship to B.L.d.B.

## Supplementary materials

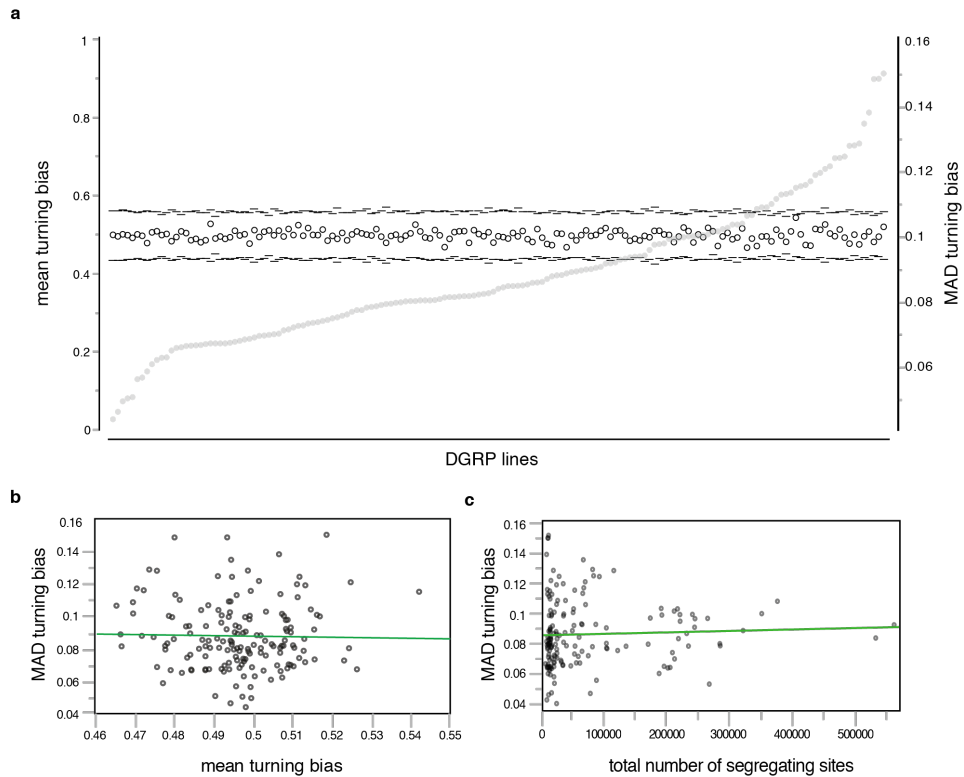

**Supplementary Figure 1 – Turn bias variability is correlated neither with the mean of turning bias nor residual genetic variation within line.** *a.* Distribution of the mean turning bias for each line, ranked by the MAD of turning bias (grey dots, scale on the right axis). Tick marks represent a 99.9% CI around the mean. There is no significant difference between lines in mean turning bias. Each lines are on average un-handed, making equal portions of left and right turns. ANOVA *p*-value < 0.87. *b.* No relationship between the mean turning bias and intragenotypic variability ($r^2$ = 0.0004, *p*-value = 0.80). *c.* No relationship between intragenotypic variability and residual genetic variation segregating within line ($r^2$ = 0.0005, *p*-value = 0.78), data from[35].

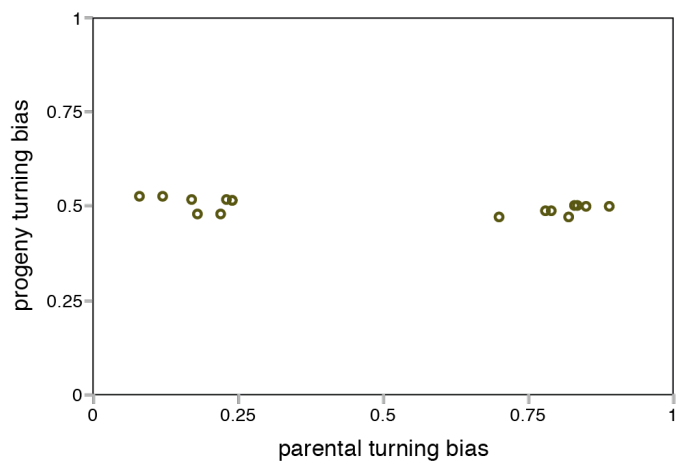

**Supplementary Figure 2 – No genetic variation segregating *within* lines is associated with turning bias.** Crosses between extreme left turning parents (at left) and crosses between extreme right turning parent (at right) produce unbiased F1 with a 50/50 left-right turning bias. Data modified from[4].

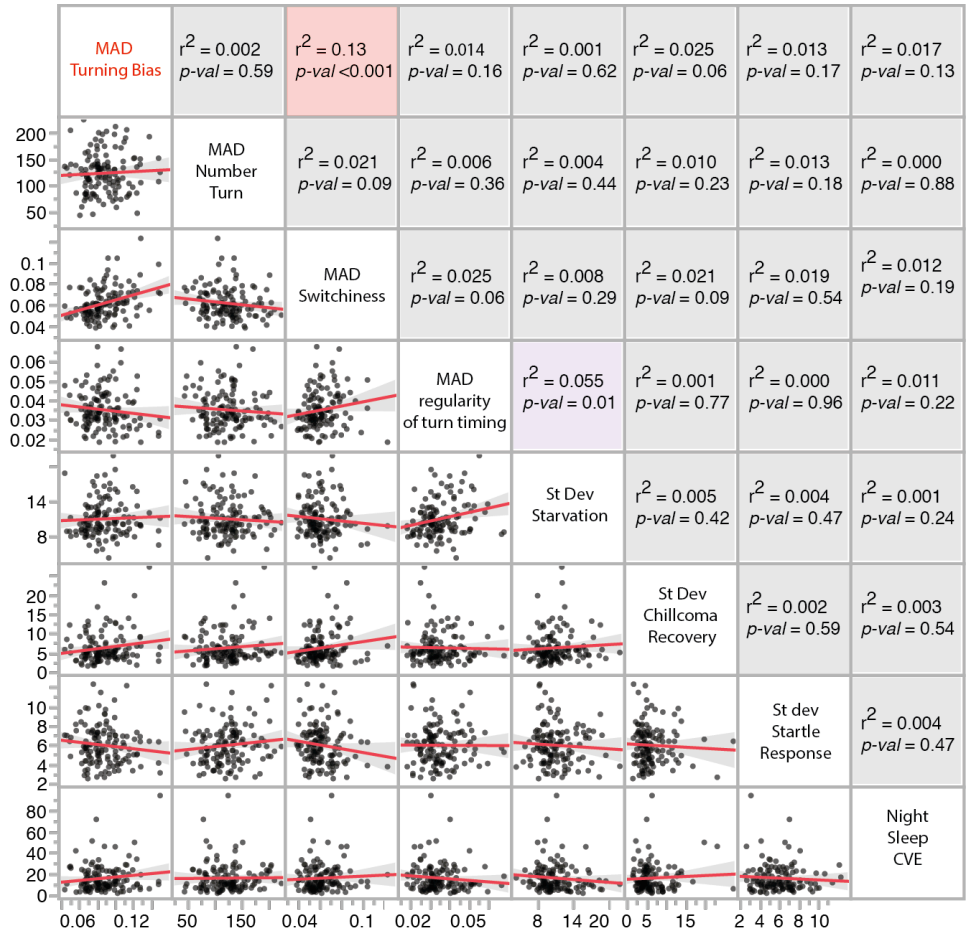

**Supplementary Figure 3 – Intragenotypic variability in various phenotypes is predominantly uncorrelated.** Scatter plots of pairs of measures of intragenotypic variability. Points are DGRP lines. Red line is linear fit with 95% confidence interval in gray. Standard deviation for starvation resistance, chill coma recovery and startle response were calculated based on data from[22]. CVE for night sleep data from[23]. For starvation resistance *n* per line = 40, chillcoma *n* per line = 50, startle response *n* per line = 40, night sleep *n* per line = 32.

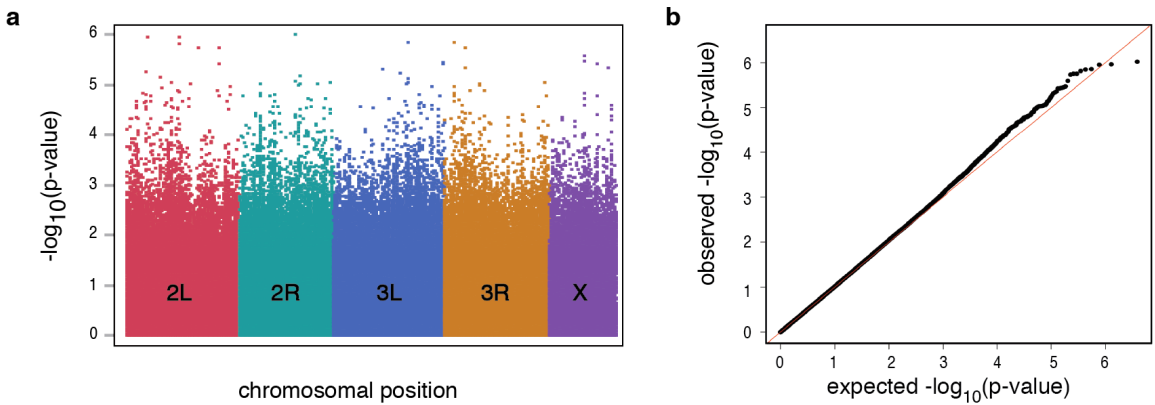

**Supplementary Figure 4 – GWAS *p*-value distributions**. *a.*–log10(*p*-value) plotted along each chromosomal position for all SNPs. Colors and letters indicate chromosome arms *b*. QQ plot comparing observed *p*-value to a uniform distribution of expected *p*-value (lambda = 1.07).

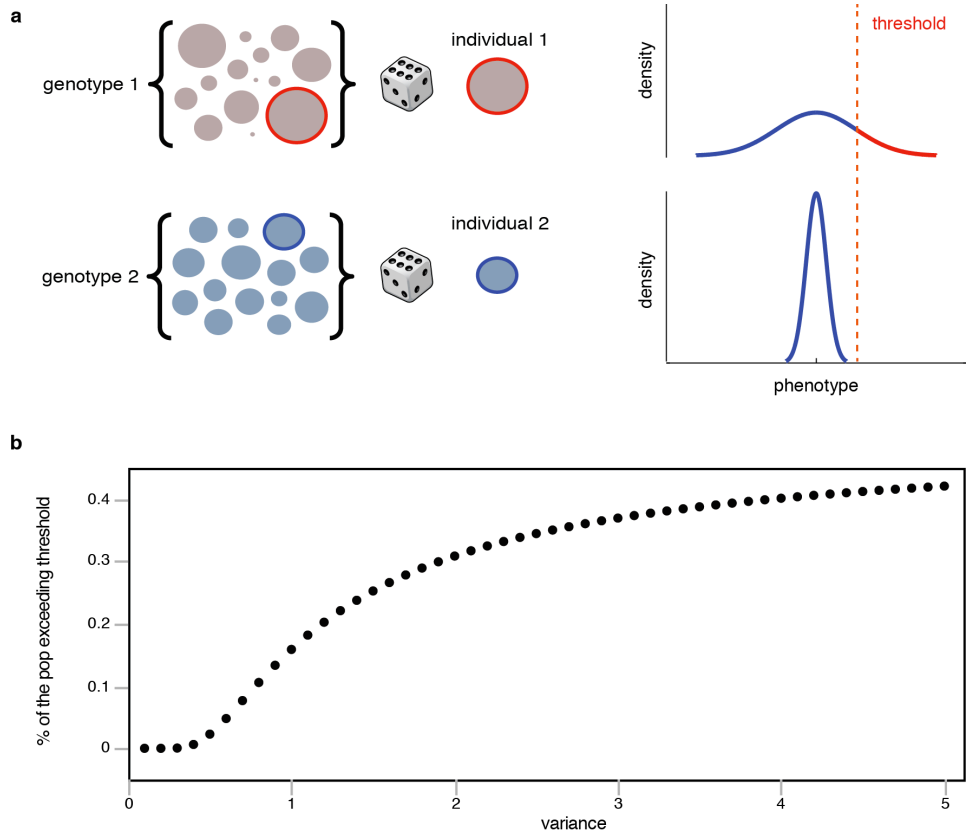

**Supplementary Figure 5 – Consequences of intragenotypic variability on the fraction of a hypothetical population exceeding a disease threshold.** *a.* Visual representation of the effects of variance on the prevalence of phenotype exceeding a threshold, such as a disease. Genotype 1 and 2 differ in their degree of intragenotypic variability. The sets of circles at left represent the range of possible outcomes for each genotype. Generally, each individual in an outbred diploid organism is a unique instance of its genotype. By contrast, our experiments with inbred lines allow us to consider multiple individuals from the same distribution. An individual drawn at random from genotype 1 (high variability) may land in the tail of the distribution, potentially in disease space. On the other hand an individual drawn randomly from genotype 2 never gets a chance to explore the phenotypic space explored by genotype 1, even if it is just as much of an outlier within its respective distribution. *b.* For a hypothetical normally-distributed population, we computed the fraction passing a disease threshold as a function of changes in variance. The distribution mean was set at 1 and the disease threshold was set at 2. It is evident that the fraction of the population passing the disease threshold strongly depends on the variability.

**Analysis of variance for mean turning bias**

| Effect | df | F | *p* |
|---|---|---|---|
| Line $_{random}$ | 158 | 0.88 | 0.85 |
| Block $_{random}$ | 28 | 1.12 | 0.29 |
| Line * Block $_{random}$ | 772 | 1 | 0.49 |
| Box | 5 | 0.41 | 0.84 |
| Maze-Array | 11 | 0.52 | 0.88 |
| Box*Maze-Array | 49 | 1.13 | 0.26 |

**Analysis of variance for the absolute median deviation of turning bias**

| Effect | df | F | *p* |
|---|---|---|---|
| Line $_{random}$ | 158 | 4.31 | <.00001 |
| Block $_{random}$ | 28 | 0.82 | 0.74 |
| Line * Block $_{random}$ | 772 | 1.04 | 0.2 |
| Box | 5 | 1.76 | 0.11 |
| Maze-Array | 11 | 0.67 | 0.76 |
| Box*Maze-Array | 49 | 1.16 | 0.22 |

**Alternative test for heterogeneity of variance between DGRP lines for turning bias**

| Test | df | F | *p* |
|---|---|---|---|
| O'Brien | 158 | 8.5953 | <.00001 |
| Brown-Forsythe | 158 | 7.567 | <.00001 |
| Levene | 158 | 7.701 | <.00001 |
| Bootstrap | Line specific results at online companion data page | | |
| ANOMV | Line specific results at online companion data page | | |

**Supplementary Table 1 – Statistics for analysis of mean and variance across DGRP lines for turning bias.** df: degrees of freedom; F: F ratio statistic; *p*: *p*-value for F ratio statistic.

| Chrs | Position | Variant | Minor Allele | Major Allele | MAF | Minor Allele Count | Major Allele Count | Mixed model Pval | Flybase ID | Gene ID | Genomic annotation |
|------|----------|---------|--------------|--------------|------|--------------------|--------------------|------------------|------------|---------|--------------------|
| 2R | 13909827 | SNP | C | T | 0.10 | 15 | 141 | 9.50E-07 | FBgn0034289 | CG10910 | UTR_3_PRIME |
| 2L | 9129764 | SNP | C | G | 0.05 | 8 | 145 | 1.09E-06 | FBgn0052982 | CG32982 | INTRON |
| 2L | 3811210 | SNP | T | C | 0.11 | 17 | 140 | 1.11E-06 | FBgn0031573 | CG3407 | SYNONYMOUS_CODING |
| 3R | 5804797 | DEL | T | TT | 0.28 | 40 | 105 | 1.39E-06 | FBgn0263097 | Glut4EF | INTRON |
| 3L | 13006450 | SNP | C | A | 0.17 | 27 | 129 | 1.42E-06 | FBgn0036333 | MICAL-like | SYNONYMOUS_CODING |
| 2L | 9121721 | SNP | T | G | 0.07 | 11 | 146 | 1.53E-06 | | | |
| 2L | 16392736 | SNP | T | C | 0.37 | 52 | 87 | 1.77E-06 | | | |
| 3R | 9419254 | SNP | C | G | 0.22 | 33 | 119 | 1.77E-06 | FBgn0038159 | CG14369 | UPSTREAM |
| 2L | 12447996 | SNP | A | T | 0.40 | 56 | 85 | 1.84E-06 | FBgn0032434 | CG5421 | INTRON |
| X | 12153076 | SNP | T | A | 0.08 | 12 | 141 | 2.54E-06 | FBgn0259240 | Ten-a | INTRON |
| X | 12153062 | SNP | A | C | 0.08 | 12 | 143 | 3.37E-06 | FBgn0259240 | Ten-a | INTRON |
| 3L | 23001316 | SNP | T | C | 0.08 | 12 | 144 | 3.54E-06 | FBgn0262509 | nrm | INTRON |
| 2L | 16392768 | SNP | G | A | 0.34 | 49 | 94 | 3.64E-06 | | | |
| 3L | 23001309 | SNP | A | G | 0.08 | 12 | 143 | 3.74E-06 | FBgn0262509 | nrm | INTRON |
| X | 16161192 | SNP | T | A | 0.35 | 54 | 100 | 3.75E-06 | FBgn0040207 | kat80 | UTR_3_PRIME |
| X | 19449711 | SNP | G | A | 0.06 | 9 | 149 | 4.38E-06 | | | |
| 3R | 9414756 | SNP | G | T | 0.25 | 38 | 112 | 4.58E-06 | FBgn0038158 | CG14370 | DOWNSTREAM |
| 3L | 8951894 | SNP | C | G | 0.22 | 35 | 121 | 4.77E-06 | FBgn0035941 | CG13313 | UPSTREAM |
| 2L | 3572163 | SNP | A | T | 0.20 | 30 | 120 | 5.26E-06 | | | |
| 3L | 11733450 | SNP | A | G | 0.20 | 30 | 118 | 5.81E-06 | FBgn0036202 | CG6024 | INTRON |
| 2R | 14651257 | DEL | AA | A | 0.20 | 30 | 122 | 6.44E-06 | FBgn0034389 | Mctp | INTRON |
| 2L | 6209462 | SNP | A | G | 0.05 | 8 | 148 | 7.03E-06 | FBgn0085409 | CG34380 | UTR_3_PRIME |
| 3L | 13006461 | SNP | A | C | 0.16 | 25 | 128 | 7.33E-06 | FBgn0036333 | MICAL-like | NON_SYNONYMOUS_CODING |
| 2R | 13909829 | SNP | A | G | 0.10 | 16 | 140 | 8.09E-06 | FBgn0034289 | CG10910 | UTR_3_PRIME |
| 2L | 8003155 | SNP | G | A | 0.18 | 24 | 110 | 8.58E-06 | FBgn0031972 | Wwox | INTRON |
| 3R | 26478454 | SNP | C | T | 0.46 | 67 | 80 | 8.69E-06 | | | |
| 3L | 12859591 | SNP | A | G | 0.17 | 27 | 128 | 8.82E-06 | | | |
| 2R | 20092066 | SNP | A | G | 0.12 | 19 | 136 | 9.08E-06 | FBgn0034990 | CG11406 | INTRON |
| 3R | 13031950 | SNP | G | T | 0.43 | 62 | 83 | 9.13E-06 | | | |
| 2R | 17929411 | SNP | G | A | 0.09 | 14 | 141 | 9.16E-06 | FBgn0005778 | PpD5 | SYNONYMOUS_CODING |
| 2L | 10873583 | DEL | T | TGA | 0.20 | 29 | 117 | 9.21E-06 | | | |
| 2R | 14823241 | SNP | A | G | 0.07 | 11 | 144 | 9.26E-06 | FBgn0034408 | sano | INTRON |
| 3L | 12688628 | SNP | T | C | 0.11 | 17 | 141 | 9.29E-06 | FBgn0014343 | mirr | INTRON |
| 2R | 7130275 | SNP | T | G | 0.17 | 26 | 129 | 9.52E-06 | FBgn0033593 | Listericin | UPSTREAM |
| 3R | 13031960 | SNP | C | A | 0.42 | 61 | 85 | 9.85E-06 | | | |
| 2R | 14548570 | SNP | T | C | 0.48 | 75 | 81 | 9.89E-06 | FBgn0259202 | CG42306 | INTRON |

**Supplementary Table 2 – Top GWAS hits for MAD of turning bias.**